# Lagrangian approach for the study of heat transfer in a nuclear reactor core using the SPH methodology

F. Pahuamba-Valdez[2], E. Mayoral-Villa[1], C. E. Alvarado-Rodríguez[3], J. Klapp[1], A. M. Gómez-Torres[1] and E. Del Valle-Gallegos[2]

[1]Instituto Nacional de Investigaciones Nucleares, Carretera México – Toluca s/n, La Marquesa, 52750 Ocoyoacac, Estado de México, México.
[2]Instituto Politécnico Nacional, Escuela Superior de Física y Matemáticas, Instituto Politécnico Nacional, Ciudad de México, México.
[3]Departamento de Ingeniería Química, División de Ciencias Naturales y Exactas, Universidad de Guanajuato Noria Alta s/n, 36050 Guanajuato, Gto. México.

**Summary**

Numerical modeling simulations and the use of high-performance computing are fundamental for detailed safety analysis, control and operation of a nuclear reactor, allowing the study and analysis of problems related with thermal-hydraulics, neutronic and the dynamic of fluids which are involved in these systems. In this work we introduce the bases for the implementation of the smoothed particle hydrodynamics (SPH) approach to analyze heat transfer in a nuclear reactor core. Heat transfer by means of convection is of great importance in many engineering applications and especially in the analysis of heat transfer in nuclear reactors. As a first approach, the natural convection in the gap (space that exists between the fuel rod and the cladding) can be analyzed helping to reduce uncertainty in such calculations that usually relies on empirical correlations while using other numerical tools. The numerical method developed in this work was validated while comparing the results obtained in previous numerical simulations and experimental data reported in the literature showing that our implementation is suitable for the study of heat transfer in nuclear reactors. Numerical simulations were done with the DualSPHysics open source code that allows to perform parallel calculations using different number of cores. The current implementation is a version written in CUDA (Compute Unified Device Architecture) that allows also the use of GPU processors (Graphics Processor Unit) to accelerate the calculations in parallel using a large number of cores contained in the GPU. This makes possible to analyze large systems using a reasonable computer time. The obtained results verified and validated our method and allowed us to have a strong solver for future applications of heat transfer in nuclear reactors fuel inside the reactor cores.

# 1 Introduction

Safety analysis in a nuclear reactor is a subject of big interest for the evaluation of operational transients and prevention of severe accidents. Due to the multi physical nature of the phenomena taking part in a nuclear reactor core, the use of numerical modeling and high-performance computing has become a must in the development of numerical tools for analysis of nuclear reactors. In a boiling water reactor (BWR) the nuclear fuel, the moderator, the control elements and part of the heat removal system are found in the same structure within the vessel of the reactor. This array presents the complexity that water is used both, as moderator (fundamental in the neutronic behavior of the reactor and totally related to power generation) while it serves as vehicle of heat removal as a coolant. When water and fuel come into contact, the latter transfers heat to the water triggering the vapor generation. The vapor generated goes through the turbines of the turbo generators transforming the enthalpy of vapor into electrical power. When the transition occurs (onset of nucleate boiling) the properties of heat transfer are affected by the nucleation and the process of boiling alters the dynamics of the fluid. In BWR reactors, the transition of the liquid phase into gas plays and important role in the design and control of the reactor. The most important thermal limits for design and operation of the boiling reactor are related with the capability of the water to continue the coolant process even though the change of phase from liquid to gas.

An important phenomenon that must be considered in the BWR, is the reaching of a critical heat flux in the cladding of the fuel in which a film of vapor can be formed in the cladding acting as a thermal isolator, damaging the heat transfer and thus, increasing dramatically the temperature in the fuel. Consequently, the fuel overheats and can melt. This phenomenon is known as '*dry-out*' and, as already stated, it is of great interest to study in detail. One of the first steps in the analysis of this phenomenon is to analyze the heat transfer in the space between the fuel rod and the inner cladding wall. This space, known as gap, is full of helium and is usually coarse approximated by heat conduction phenomenon in current numerical tools. Since heat transfer in such gap plays an important role in the heat transfer from fuel to coolant outside of the cladding wall, it is imperative to improve the numerical models in the gap. For that, the natural convection model developed, verified and validated in this paper becomes very important.

Furthermore, to analyze the flow pattern of the coolant on the other side of the cladding wall is also a key aspect in nuclear reactor analysis. The turbulent behavior of the fluid, mostly in the region of heat transfer favors the transfer of heat in the system and its control is important to optimize its proper functioning. For this reason, special designs can be made to induce turbulence that can favor, along with the boiling processes, the extraction of heat generated by the fission processes. This phenomenon will be studied in detail in a future development to extend the current one.

In general, to study these systems, Eulerian methods have been traditionally been used such as finite elements, finite volumes and finite differences. These type of methods and Computational Fluid Dynamics (CFD) programs running in a supercomputer have demonstrated to be an important tool in the design, control and operation of PWR reactors which operates at high pressure and thus without boiling. Nevertheless, for BWR's, due to the transition phase, its application presents some noteworthy restrictions and limits, especially while dealing with mobile and diffused interphases, phase transitions, complex geometry and dynamic systems or turbulent fluids, which are cases that represent many of the fundamental phenomena that take place in a BWR reactor. The main limitation of Eulerian solutions is that the systems are solved by the discretization of coupled differential equations that solve the system in a mesh which is not dynamic or adaptive to the structural changes that are present, so the use of approximations must be applied. At the same time the incorporation of others physical and chemical effects results in an increase of the complexity of the numeric solution that can carry strong numeric instabilities. The traditional techniques for one-dimensional models are based on simplified models where empirical correlations and approximations are incorporated introducing uncertainties hard to quantify. Therefore, it is necessary to incorporate physical models of first principles to represent the main phenomena that occur in the boiling water reactors such as boiling and condensation, mass and energy exchange between phases, transport of particles, etc. in substitution of the empirical correlations that additionally have intervals of limited validity.

To overcome these difficulties, the use of a free mesh simulation like the SPH method has been considered as a promissory option. SPH is a Lagrangian methodology and is based on interpolation theory that uses points, generally called particles, to discretize the continuous medium. The SPH is a computational method used to simulate the dynamics of continuous mediums such as the mechanics of solids and the flux of fluids. Initially, it was developed by Gingold and Monaghan [1] for problems related to astrophysics. Its use expands over several fields of investigation including astrophysics, ballistics, volcanology and oceanography. The equations in the medium or continuous fluid which are the conservation of mass, moment and energy of the fluid, are present in Lagrangian form and later discretized using the SPH methodology.

The numerical simulations were performed with the DualSPHysics free code (for details please refer to [5] and *www.dual.sphysics.org*). One of the main advantages of the DualSPHysics code is its parallel structure. The code is written in the C++ language using the Open Multi-processing application that allows to perform calculations in parallel using different number of cores according to the computer equipment used. In addition, there is a version written in CUDA (Compute Unified Device Architecture) that allows to use the GPU processors (Graphics Processor Unit) to accelerate the calculations in parallel using many cores contained in the GPU. The ability of GPUs to perform numerical simulations using the SPH method is demonstrated by Harada [7] where a speedup of 28 was achieved by simulating 60,000 particles. The DualSPHysics code is organized mainly in three stages: (1) the creation of the list of neighbors, (2) the computation of the interaction between neighbors, (3) the integration in time, referring to the update of the system. Crespo et al. [5] verified that the interaction stage is the one that consumes the most computation time in a numerical simulation. Based on the above, to improve the calculation performance it is necessary to perform the interaction stage in parallel, in this way the sequential calculation is avoided, and the interaction of several particles is calculated at the same time using different cores.

In the second part of this work the general characteristics of the SPH approach are presented together with the considerations needed to introduce the bases for the implementation to analyze heat transfer in the fuel of nuclear reactors. In the third section the results obtained for the study of heat convection, which is of great importance in many applications of engineering and especially in the analysis of heat transfer in nuclear reactors, are presented. The 2D models developed here will be extended to 3D models to study the natural convection in the gap (space that exists between the fuel rod and the cladding) and the comparison with previous numerical simulations. Section 4 contains the conclusions of this work.

## 2 SPH methodology

The generalities of the SPH approach are described in this section. The equations that govern the dynamics of the continuous media (fluids and deformable solids) are transformed into integral equations through the use of an interpolation function. Thus, in SPH the medium is represented numerically by a finite set of observation points, or particles, by means of a smoothing procedure in which the estimated value of a function $f(x)$ at a point $x$ is given by the expression

$$f(x) = \int_{\Omega} f(x')W(x - x', h)dx', \tag{1}$$

where $W(x - x', h)$ is the smoothing function usually called kernel, which is a function of the position $x$ and a smoothing length $h$ that determines the domain of influence $\Omega$.

If a fluid of density $\rho(x)$ is considered, the interpolation integral shown in equation (1) can be written as

$$\int_{\Omega}\left[\frac{f(x')}{\rho(x')}\right]W(x-x',h)\rho(x')dx'. \tag{2}$$

To evaluate this integral the domain $\Omega$ is subdivided into $N$ elements of volume (particles) each of mass $m_b$ and density $\rho_b$, in such a way that the sum of the masses of all the particles is the total mass of the fluid. Thus, the mass of each particle ($m_b$) can be identified as

$$m_b = \rho(x')dx', \tag{3}$$

where $dx'$ is the volume differential and $\rho(x')$ is the density.

Equation 2 can be discretized in a set of particles by replacing the interpolation integral with the summation over the mass of the particle $m_b$

$$f(x) = \sum_{b=1}^{N} m_b \frac{f_b}{\rho_b} W(|x-x_b|, h), \tag{4}$$

where the subscript $b$ refers to the quantity evaluated in position $b$.

An advantage of the SPH method is that its formulation allows the first derivative to be estimated in a simple way considering the kernel as a differentiable function obtaining:

$$\frac{\partial f(x)}{\partial x} = \sum_{b=1}^{N} m_b \frac{f_b}{\rho_b} \frac{\partial W(|x-x_b|, h)}{\partial x}. \tag{5}$$

In this way, in SPH each derivative is calculated from the exact derivative of the kernel function. This feature allows to calculate the gradient of any function $f(r)$ in a simple manner through the kernel gradient in such a way that

$$\nabla f(r) = \sum_{b=1}^{N} m_b \frac{f_b}{\rho_b} \nabla W(|r_a - r_b|, h) = \sum_{b=1}^{N} m_b \frac{f_b}{\rho_b} \nabla_a W_{ab}, \tag{6}$$

where $r$ is the position vector.

A graphic description of the kernel function is shown in Figure 1. The smoothing length is usually constant, however there are works that report algorithms to use a variable $h$ value in each particle [2, 3].

The kernel function can be written as follows on the position $r$ and the smoothing length $h$

$$W(r,h) = \frac{\sigma}{h^v} f(q), \tag{7}$$

where $q = r/h$ and $v$ is the number of spatial dimensions.

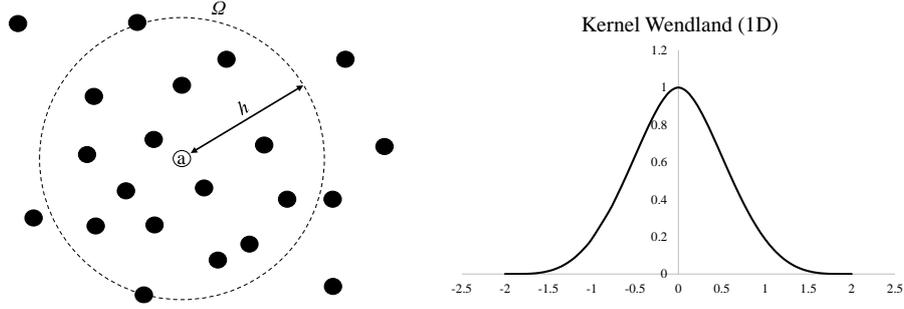

Figure 1. a) Representation of the particle of interest "a", neighboring particles (black points), smoothing length ($h$) and domain of interest ($\Omega$) in the kernel function, b) example of a kernel function (Wendland) in 1D.

The normalization condition is expressed as:

$$\sigma \int f(q) dV = 1,$$

where $dV = dq$, $2\pi q dq$ and $4\pi q^2 \, dq$ in 1D, 2D and 3D, respectively.

### 2.1 SPH formalism used for the equations of fluid dynamics.

In this section are described the equations that govern the fluid dynamics in SPH formalism which are obtained from the continuous form of each equation using the methodology shown in the interpolation section.

The equation of continuity described in Lagrangian form is presented in equation (8) as

$$\frac{d\rho}{dt} = -\rho \nabla \cdot v. \tag{8}$$

For particle *a* we have:

$$\frac{d\rho_a}{dt} = -\rho_a (\nabla \cdot v)_a. \tag{9}$$

Using equation (6) to evaluate the gradient, it is obtained:

$$\frac{d\rho}{dt} = -\rho_a \sum_{b=1}^{N} m_b \frac{v_b}{\rho_b} \cdot \nabla_a W_{ab}. \qquad (10)$$

The momentum conservation equation is defined as

$$\frac{dv}{dt} = \frac{-1}{\rho}\nabla P + g + v_0 \nabla^2 v + \frac{1}{\rho}\nabla \cdot \tau, \qquad (11)$$

where the laminar term ($v_0 \nabla^2 v$) is treated as in equation (12) and $\tau$ represents the stress tensor. The wall shear stress tensor is calculated from the SPH model according to the equation (12) that is completely described in [5].

$$\frac{dv_a}{dt} = -\sum_b m_b \left(\frac{P_b}{\rho_b^2} + \frac{P_a}{\rho_a^2}\right) \nabla_a W_{ab} + g$$

$$\sum_b m_b \left(\frac{4v_0 r_{ab} \cdot \nabla_a W_{ab}}{(\rho_a + \rho_b)(r_{ab}^2 + \eta^2)}\right) v_{ab} +$$

$$\sum_b m_b \left(\frac{\tau_{ij}^b}{\rho_b^2} + \frac{\tau_{ij}^a}{\rho_a^2}\right) \nabla_a W_{ab} \qquad (12)$$

*2.2. SPH Model for Natural Convection*

For the numerical simulation the conservation equations of momentum, mass and energy in Lagrangian formalism are considered:

$$\frac{d\rho}{dt} = -\rho \nabla \cdot v, \qquad (13)$$

$$\frac{dv}{dt} = \frac{-1}{\rho}\nabla P + \frac{\mu}{\rho}\nabla^2 v + F^B, \qquad (14)$$

$$\frac{dT}{dt} = \frac{1}{\rho C_p} \nabla \cdot (k \nabla T), \qquad (15)$$

where $\rho$ is the density, $t$ is the time, **v** is the velocity vector, $P$ is the pressure, $\mu$ is the viscosity, $\boldsymbol{F^B}$ is the buoyant force, $T$ is the temperature, $C_p$ is the heat capacity and $k$ is the thermal conductivity coefficient.

The motion of the fluid due to the change in temperature is provided by Boussinesq approximation:

$$F^B = -g\beta(T - T_r) , \qquad (16)$$

where $g$ is the gravitational acceleration vector, $\beta$ is the thermal coefficient of volumetric expansion, $T$ is the temperature of the fluid and $T_r$ is the reference temperature of the fluid.

The momentum, continuity and energy equations can be discretized using the SPH formalism and gives:

$$\frac{d\rho}{dt} = -\rho_a \sum_{b=1}^{N} m_b \frac{v_b}{\rho_b} \cdot \nabla_a W_{ab} , \qquad (17)$$

$$\frac{dv_a}{dt} = -\sum_{b=1}^{N} m_b \left(\frac{P_b}{\rho_b^2} + \frac{P_a}{\rho_a^2} + \Gamma\right) \nabla_a W_{ab} + g , \qquad (18)$$

$$\frac{dT_a}{dt} = \frac{1}{Cp} \sum_{b=1}^{N} \frac{m_b(k_a + k_b)(r_a - r_b) \cdot \nabla_a W_{ab}}{\rho_a \rho_b (r_{ab}^2 + \eta)} (T_a - T_b) . \qquad (19)$$

Equations (17) - (18) are coupled by the Tait state equation

$$P = B\left[\left(\frac{\rho}{\rho_r}\right)^\gamma - 1\right], \qquad (20)$$

where P is the pressure, $\rho$ is the density of the fluid, $\rho_r$ is the reference density, $B = c_0^2 \rho_r / \gamma$, $\gamma = 7$ for liquids and $\gamma = 1.4$ for gases.

To consider a change in the reference density $\rho_r$ in equation (20) due to the temperature change, the following model is used by the coefficient of volumetric expansion.

$$V_f = V_0[1 + \beta(T_f - T_0)] , \qquad (21)$$

where $V_f$ and $V_0$ are the final and initial volumes respectively and $T_f$ and $T_0$ are the final and initial temperature, respectively. Relating density, mass and volume

$$\rho = \frac{m}{V}, \tag{22}$$

and substituting the volume from (22) in equation (21):

$$\rho_f = \rho\left(\frac{1}{1 + \beta(T_f - T_0)}\right), \tag{23}$$

in this way the reference density $\rho_r = \rho_f$ is evolved in equation (23) at each time step, thus the fluid tends to the value of a new density when the fluid temperature changes. This calculation is performed per particle at each time step.

Finally, in equation (24) the value of the coefficient of thermal conductivity per particle is calculated by the following expression

$$k_a = \alpha \rho_f C p. \tag{24}$$

where α is the thermal diffusivity coefficient. With the previous model the change of temperature and density of the fluid affects the coefficient of thermal conductivity considering a more robust model in comparison with the models that consider constant *k*.

*2.3 The integration algorithm, (Verlet).*

The values of position, density and speed are updated every time step using the Verlet algorithm [6], which does not require multiple calculations for each time step and has a lower computational load compared to other integration techniques. The Verlet algorithm consists of two parts: in the first part, the integration of position, density and velocity is carried out using equations (25) - (28). The second option of equations (25)-(27) is applied every certain number of steps ($t_s \approx$ 50). The second option of the algorithm prevents the results from diverging from the correct solution over time.

$$v_a^{t+1} = v_a^{t-1} + 2\Delta t \left(\frac{dv_a}{dt}\right)^t \text{ each } t_s \approx 50 \quad v_a^{t+1} = v_a^t + \Delta t \left(\frac{dv_a}{dt}\right)^t, \tag{25}$$

$$\rho_a^{t+1} = \rho_a^{t-1} + 2\Delta t \left(\frac{d\rho_a}{dt}\right)^t \text{ each } t_s \approx 50 \quad \rho_a^{t+1} = \rho_a^t + \Delta t \left(\frac{d\rho_a}{dt}\right)^t \tag{26}$$

$$T_a{}^{t+1} = T_a{}^{t-1} + 2\Delta t \left(\frac{dT_a}{dt}\right)^t \quad \text{each } t_s \approx 50 \quad T_a{}^{t+1} = T_a{}^{t-1} + \tag{27}$$

$$2\Delta t \left(\frac{dT_a}{dt}\right)^t$$

$$r_a{}^{t+1} = r_a{}^t + \Delta t v_a{}^t + 0.5\Delta t^2 \left(\frac{dv_a}{dt}\right)^t, \tag{28}$$

Each time step ($\Delta t$) reported in equations (25) - (28) is calculated using equation (31) to establish a time step that ensures stability in the simulation. Equation (31) is calculated from equations (29) and (30). Equation (29) is calculated based on the maximum acceleration in the fluid, that is, the value of the particle with the greatest acceleration is considered. Equation (30) considers the speed of sound ($c_s = c_0\rho^3$) as well as the value of the viscosity ν. In addition, the passage of time is controlled using the Courant-Friedrich-Levy (CFL) condition [4],

$$\Delta t_f = \min_a \left(\sqrt{h/\left|\frac{dv_a}{dt}\right|}\right), \tag{29}$$

$$\Delta t_{cv} = \min_a \frac{h}{c_s + \max|hv_{ij}r_{ij}/(r^2{}_{ij}+\eta^2)|}, \tag{30}$$

$$\Delta t = CFL \cdot \min(\Delta t_f, \Delta t_{cv}), \tag{31}$$

where $h$ is the smoothing length, $v_{ij} = v_i - v_j$ and $\eta^2 = 0.01h^2$.

## 3 RESULTS

For the verification and validation of the mathematical modelling, two concentric tubes were simulated according to the results shown by Yang & Kong [8], in which the same ratio of $L/D_i = 0.8$ is considered according to Figure 2, where in (b) we present the initial conditions for the validation study cases. In all cases the temperature of the contours is maintained constant and the only change is the relationship that exists in the dimensionless numbers of Rayleigh (*Ra*) and Prandtl (*Pr*), specifically the value of the thermal diffusivity of the fluid (α).

$$Ra = \frac{g\beta L^3 \Delta T}{v\alpha}; \quad Pr = \frac{v}{\alpha}.$$

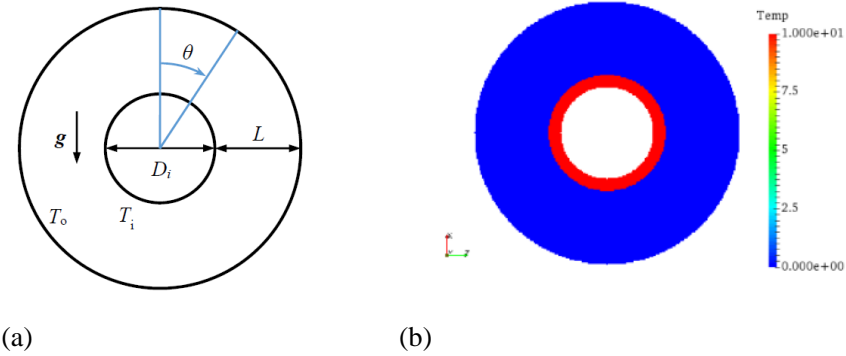

(a) (b)

Figure 2. (a) Dimensions of the concentric tube system and (b) Initial conditions for SPH simulations.

Numerical simulations for different cases were performed, for Ra = $10^2$ and Pr = 10, Ra = $10^4$ and Pr = 10, Ra = $10^6$ and Pr = 10. In all cases the steady state was reached and then the isothermal profiles between the concentric tubes were obtained. Figure 3 shows these results.

These results correspond well with the data reported in [8]. Validation was done also comparing with experimental data presented in [9]. For this case, Figure 4 shows the experimental system studied, which consists of a system of concentric cylinders aligned horizontally, where the two cylinders are at constant temperature of different magnitude, the inner cylinder being the highest temperature, as well as the components used to maintain the experimental conditions. The external diameter of the inner cylinder is 3.56 cm, with a thickness of 0.51 cm and the inner diameter of the outer cylinder is 9.25 cm, with a thickness of 0.45 cm, which maintains a relation $L/D_i = 0.8$.

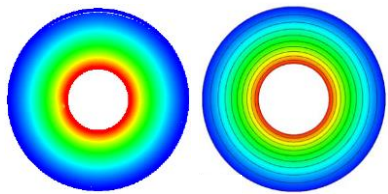

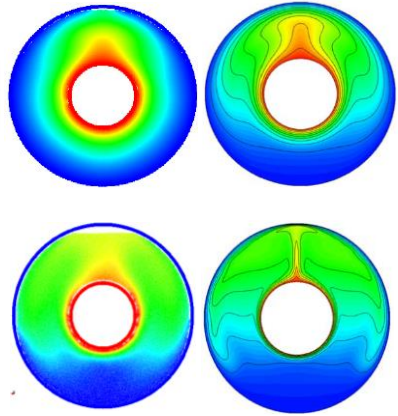

Figure 3. Comparison of the isothermal profile results with DualSPHysics and those reported by Yang & Kong [8]. From top to bottom we present the cases for $Ra = 10^2$ and $Pr = 10$, $Ra = 10^4$ and $Pr = 10$, and $Ra = 10^6$ and $Pr = 10$, respectively.

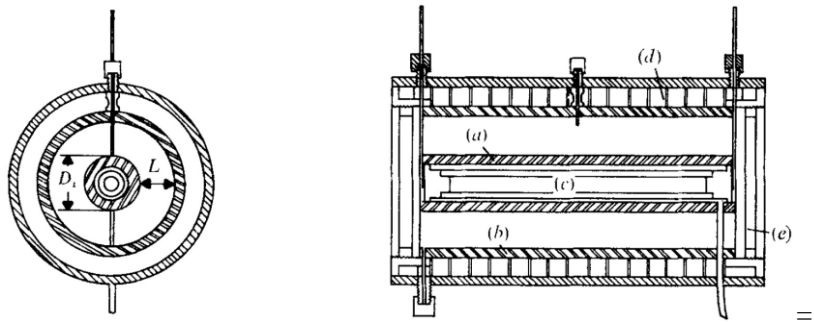

Figure 4: Diagram of the experimental facility. (a) Internal cylinder. (b) External cylinder. (c) Heater. (d) Cooling water channels. (e) Window. (Taken from reference [9]).

Figure 5 shows the temperature profiles for one of the water tests above, $Ra = 10^5$ taken from reference [9]. The thermic limit layers near both cylinders are well defined, as is the temperature inversion in the central region [9]. It should be noted that for these results, the experiment was allowed to reach a steady state, it took at least 8 hours for the circulation of the water.

As it can be seen in Figure 5, the performed calculations agree in a comprehensive manner with the experimental data. The analysis of the numerical results is shown in the Figures 6-9 where distribution of temperature, density, thermal conductivity and velocity are shown. In the Table 1 the values used in the simulation are reported.

In Figure 6, the effect that the buoyant force exerts on the fluid is appreciated, when forming what colloquially it is denominated like "plume". What shows a steady state with a 1 plume, since four different convection states are identified in numerical simulations, state stable with 1 plume (SP1), unstable state with 1 plume (UP1), stable state with n (n> 1) plumes (SPN) and unstable state with n (n> 1) plumes (UPN). In Figure 5 a comparison is made between the results of the simulation carried out in the DualSPHysics code and the results obtained were published in the reference experiment [9].

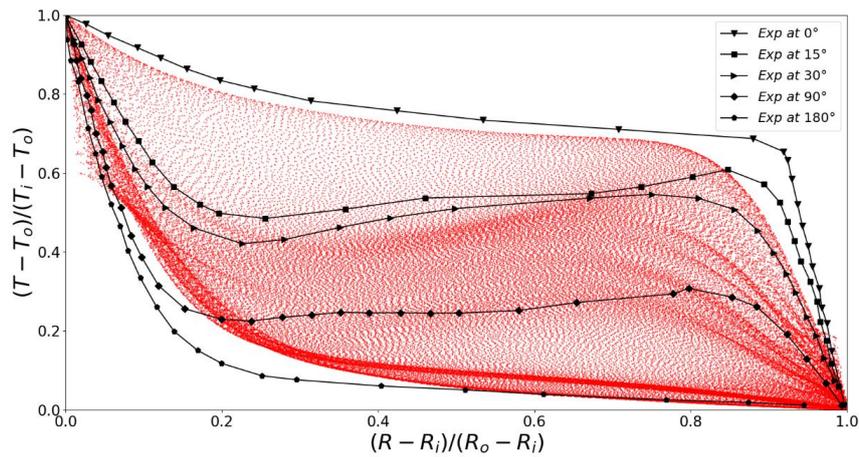

Figure 5: Dimensionless distribution of radial temperature in water for $Ra = 2.09 \times 10^5$, $Pr = 5.45$, $L/Di = 0.8$. The profiles for $\theta = 0°$, $15°$, $30°$, $90°$ and $180°$ were taken from [9] and compare with the SPH numerical results (red area).

Finally, another model was studied. In this case, a two-dimensional case with analytical solution, which consists on a plate with dimensions $L = H = 10$ cm and boundary conditions of constant temperature, $T_1 = 0 °$ C. The fluid is at an initial temperature, $T_0 = 100 °C$ [10]. Figure 10 shows the problem to be solved, with the spatial domain established together with the boundary conditions and the initial conditions.

Table 1. Parameters used in the simulation of concentric tubes.

| Parameter | Value |
| --- | --- |
| Initial distance between particles | 0.02 cm |
| Viscosity (Laminar Viscosity Treatment + SPS) | $1 \times 10^{-6}$ m$^2$/s |
| Initial temperature of the fluid | 278.15 K |
| Step Algorithm | Verlet |
| Kernel | Wendland |
| Simulation time | 3.15 seconds |
| Temperature of the cold boundary | 278.15 K |
| Temperature of the hot boundary | 348.15 K |
| Specific heat capacity at constant pressure (Cp) | 4.1813 kJ/kgK |
| Thermal diffusivity coefficient ($\alpha$) | $1.84 \times 10^{-7}$ m$^2$/s |
| Volumetric expansion coefficient ($\beta$) | $5.82 \times 10^{-4}$ °C$^{-1}$ |
| Boundary particles | 17644 |
| Fluid particles | 143192 |

In reference [10] the problem was simulated with a number of particles SPH, N = 1600, as shown in Figure 11. The numerical results obtained for different instants of time are presented in Figure 12.

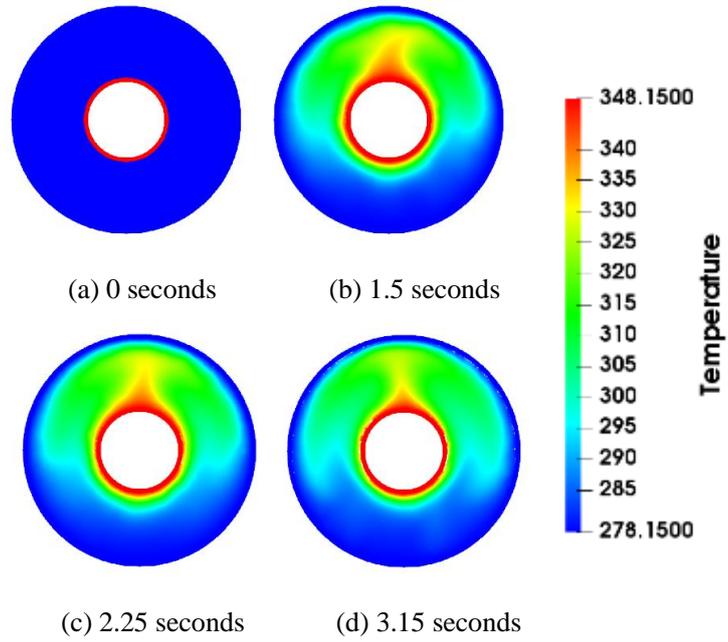

Figure 6: Distribution of the temperature at different simulation times.

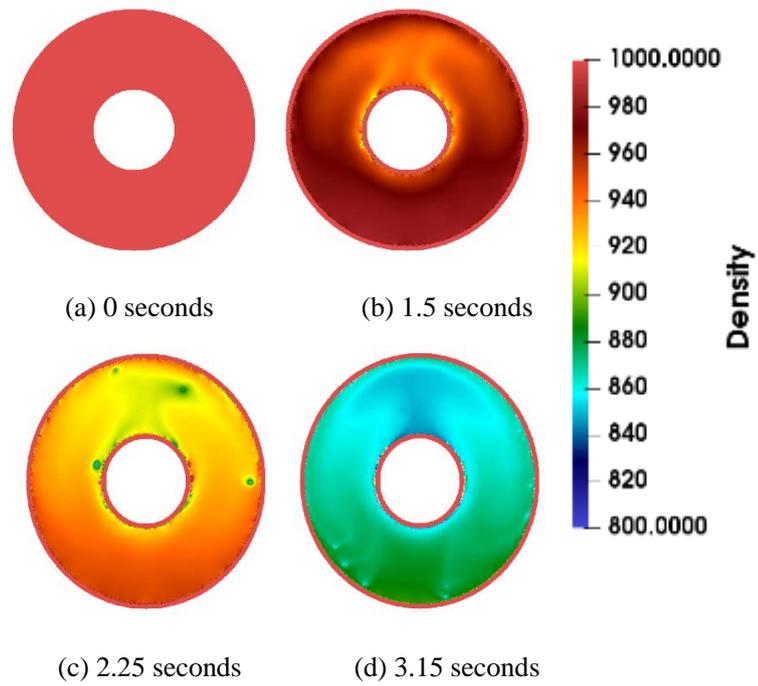

Figure 7: Variation of the density at different simulation times.

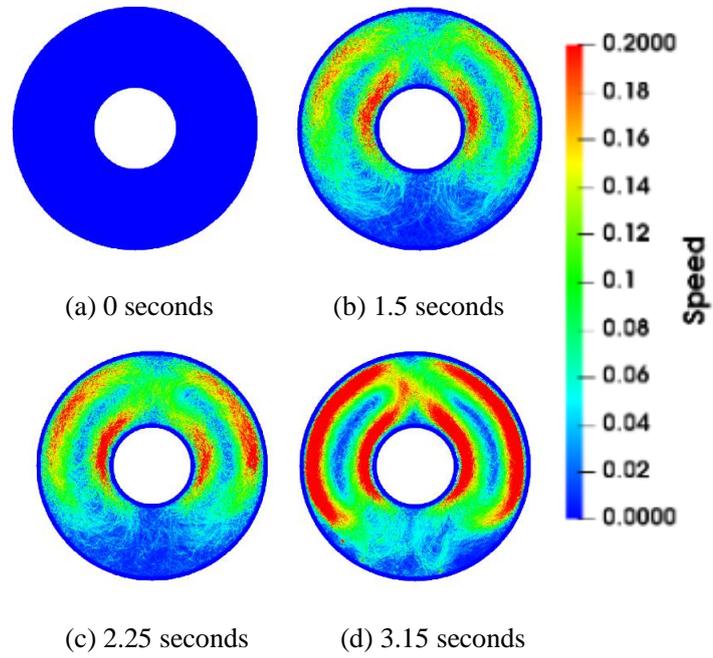

(a) 0 seconds  (b) 1.5 seconds

(c) 2.25 seconds  (d) 3.15 seconds

Figure 8: Speed profiles at different simulation times.

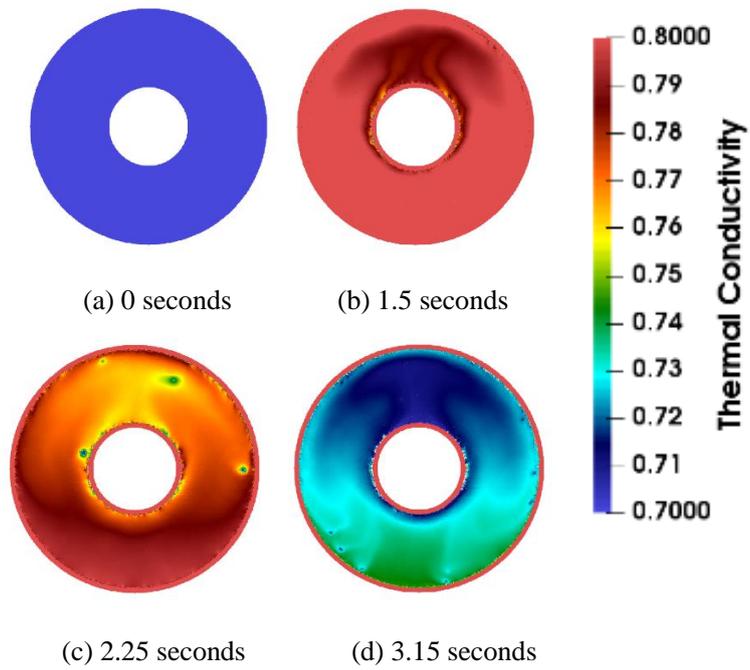

(a) 0 seconds  (b) 1.5 seconds

(c) 2.25 seconds  (d) 3.15 seconds

Figure 9: Change of thermal conductivity at different simulation times.

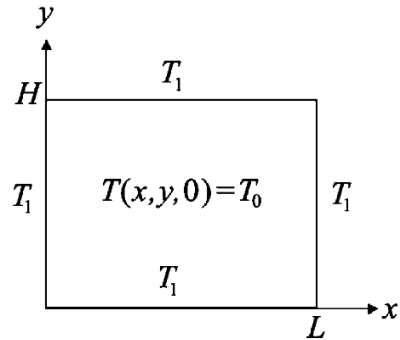

Figure 10. 2D spatial domain with boundary conditions of constant temperature and initial conditions (Taken from [10]).

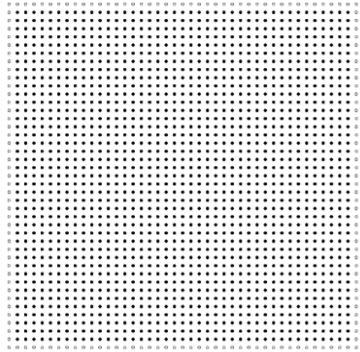

Figure 11. Spatial discretization of the 2D spatial domain using SPH particles.

Table 2 shows the parameters used in the second simulation and the Figure 12 shows different time instants of the simulation, which show the temperature distribution in the established domain.

Table 2: Simulation parameters used in the second simulation.

| Parameter | Value |
| --- | --- |
| Initial distance between particles | 0.25 cm |
| Viscosity (Laminar Viscosity Treatment + SPS) | $1 \times 10^{-6}$ m$^2$/s |
| Initial temperature of the fluid | 373.15 K |
| StepAlgorithm | Verlet |
| Kernel | Wendland |
| Simulation time | 8 seconds |
| Temperature of the boundary | 273.15 K |
| Specific heat capacity at constant pressure | 4.1813 kJ/kgK |
| Thermal diffusivity coefficient | $1.0 \times 10^{-4}$ m$^2$/s |
| Volumetric expansion coefficient | $6.95 \times 10^{-4}$ °C$^{-1}$ |
| Boundary particles | 81 |
| Fluid particles | 1600 |

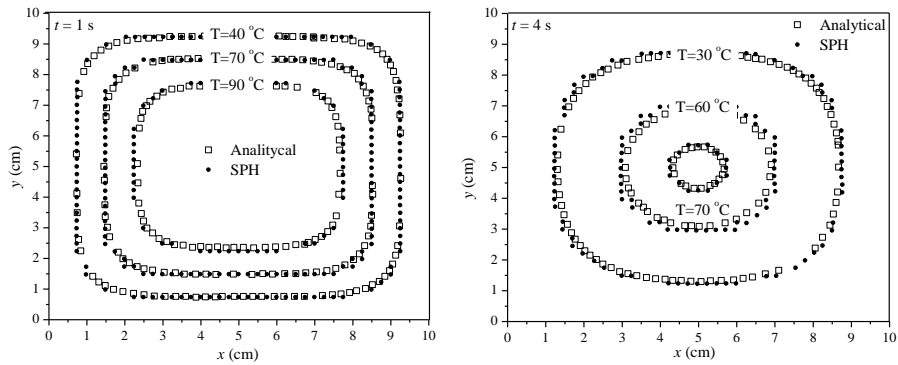

Figure 12. Comparison of isothermal contours for the SPH numerical results and the analytical solution for $t = 1.0$ and $t = 4.0$ seconds [10].

The final test show that the mathematical modelling and its solution using the SPH method presents good accuracy for simple cases with analytical solution. Moreover, when the variable thermal conductivity is applied in the mathematical modelling, the accuracy increases for values at the center of the solution comparing whit the numerical results shown by the reference [10].

**4 CONCLUSIONS**

Despite being a relatively new method for the solution of fluid dynamics problems, the SPH method is more flexible and versatile than mesh-based methods. It has been shown that with the developed tool it is possible to model the heat transfer under natural convection between two concentric rings like the phenomenon of heat transfer between fuel rod and inner cladding wall (heat transfer in fuel gap), which is a fundamental step in safety analysis of nuclear reactors related to thermal limits. Once the verification and validation of the model is finished, next step is to expand to a 3D version of the code in order to be able to perform practical analysis and comparisons with other numerical models that solve in traditional way the heat transfer in the fuel elements of a nuclear reactor.

## Acknowledgements

The authors are grateful for the financial support received from the strategic project No. 212602 (AZTLAN Platform) of the Energy Sustainability Sector Fund CONACyT-SENER for the elaboration of this work. Likewise, the author Felipe de Jesus Pahuamba Valdez thanks the National Polytechnic Institute and CONACyT for the scholarship received for their master's studies and Dr. Carlos Enrique Alvarado Rodriguez, for the technical assistance provided for the realization of this project. We acknowledge funding from the European Union's Horizon 2020 Programme under the ENERXICO Project, grant agreement No. 828947 and under the Mexican CONACYT- SENER-Hidrocarburos grant agreement No. B-S-69926. The authors thank ABACUS: Laboratory of Applied Mathematics and High-Performance Computing of the Mathematics Department of CINVESTAV-IPN for providing the computer facilities to accomplish this work.